\documentclass[12pt]{article}

\usepackage{scicite}
\usepackage{graphicx}

\usepackage{times}

\newenvironment{sciabstract}{%
\begin{quote} \bf}
{\end{quote}}

\newcommand{\be}{\begin{equation}}
\newcommand{\ee}{\end{equation}}

\newcommand{\ifm}[1]{\relax\ifmmode#1\else$\mathsurround=0pt #1$\fi}
\newcommand{\kms}{\ifmmode\,{\rm km}\,{\rm s}^{-1}\else km$\,$s$^{-1}$\fi}
\newcommand{\kpc}{\ifmmode\,{\rm kpc}\else kpc\fi}
\newcommand{\Mpc}{\ifmmode\,{\rm Mpc}\else kpc\fi}

\newcommand{\ltsima}{$\; \buildrel < \over \sim \;$}
\newcommand{\lsim}{\lower.5ex\hbox{\ltsima}}
\newcommand{\gtsima}{$\; \buildrel > \over \sim \;$}
\newcommand{\gsim}{\lower.5ex\hbox{\gtsima}}

\newcounter{lastnote}

\title{A Cosmic Microwave Background feature consistent with a cosmic texture}

\author
{M. Cruz,$^{1,2\ast}$ N. Turok,$^{3}$ P. Vielva,$^{1}$ E. Mart\'{\i}nez-Gonz\'alez$^1$ \& M. Hobson$^4$\\
\\
\normalsize{$^{1}$IFCA, CSIC-Univ. de Cantabria,  Avda. los Castros, s/n, 39005-Santander, Spain.}\\
\normalsize{$^{2}$Dpto. de F\'{\i}sica Moderna, Univ. de Cantabria, Avda. los Castros, s/n, 39005-Santander, Spain.}\\
\normalsize{$^{3}$DAMTP, CMS, Wilberforce Road, Cambridge, CB3 0WA, UK.}\\
\normalsize{$^{4}$Astrophysics Group, Cavendish Laboratory, J.J. Thomson Avenue, Cambridge CB3 OHE, UK.}\\
\\
\normalsize{$^\ast$E-mail: cruz@ifca.unican.es.}
\\
\\
\normalsize{Accepted by Science. Published electronically via Science Express.}
\\
\normalsize{http://www.sciencemag.org/cgi/content/abstract/1148694}
}

\date{}

\begin{document} 


\baselineskip24pt


\maketitle


\begin{sciabstract}
The Cosmic Microwave Background provides our most ancient image
of the Universe and our best tool for studying its early evolution.
Theories of high energy physics predict the formation of various
types of topological defects in the very early universe, including
cosmic texture which would generate hot and cold spots in the Cosmic 
Microwave Background. We show through a Bayesian statistical analysis that the most prominent, $5^\circ$ radius cold spot observed in all-sky images, which is otherwise hard to explain, is compatible with having being caused by a texture. From this model, we 
constrain the fundamental symmetry breaking energy scale to be 
$\phi_0 \approx 8.7 \times 10^{15}$ GeV. If confirmed, this detection of a cosmic defect will probe physics at energies exceeding any conceivable terrestrial experiment.
\end{sciabstract}


\section*{}

The Cosmic Microwave Background (CMB) radiation was emitted from the hot plasma of the early universe roughly $14$ billion years ago. All-sky, multi-frequency maps of the CMB sky made by the Wilkinson Microwave Anisotropy Probe (WMAP) ({\it 1,2\/}) reveal Gaussian temperature anisotropies of the form expected in standard cosmological scenarios, tracing density variations of a part in a hundred thousand in the primordial cosmos ({\it 3\/}). 
However, several apparent anomalies in the expected Gaussian, isotropic statistical distribution have also been found ({\it 4--10\/}).
One of the most striking is a large cold spot centred on Galactic coordinates $b = -57^\circ, l = 209^\circ$, with a radius of $\approx 5 ^\circ$ ({\it 10--13\/}). 
It was detected using the Spherical Mexican Hat Wavelet (SMHW), an optimal tool for enhancing such features, and has a flat frequency spectrum, inconsistent with either Galactic foregrounds or the Sunyaev-Zel'dovich effect ({\it 13\/}). A conservative estimate of the probability of finding such a feature in Gaussian simulations, taking the effect of a posteriori selection into account, is only 1.85\%. Several radical explanations, such as huge voids or an anisotropic cosmology, have already been proposed: many have been ruled out by other cosmological observations ({\it 14--17\/}). 

Here we consider the possibility that the spot was caused by a cosmic texture ({\it 18\/}), a type of cosmic defect predicting spots in the CMB ({\it 19\/}). Cosmic defects are hypothetical remnants of symmetry-breaking phase transitions in the early universe, predicted by certain unified theories of elementary particle physics. According to these theories, the different species of elementary particle are indistinguishable in the hot early universe. As the universe cools, the symmetry between them breaks, in a phase transition analogous to the freezing of water. Just as misalignments in the crystalline structure of ice lead to defects, misalignments in the symmetry breaking in unified theories lead to the formation of cosmic defects ({\it 20,21\/}). 
Breaking a discrete symmetry produces domain walls and breaking a circle (or $U(1)$) symmetry produces cosmic strings. Textures form when a simple Lie group (like the special unitary goup, $SU(2)$) is broken. They consist of localized, twisted configurations of fields which collapse and unwind on progressively larger scales, a scaling process continuing into the late universe. Each unwinding texture creates a concentration of stress-energy and a time-varying gravitational potential. CMB photons passing through such a region receive a red- or blue-shift, creating a cold or hot spot ({\it 19\/}) with a magnitude set by the symmetry-breaking energy scale $\phi_0$. 
We have used high resolution texture simulations and Bayesian statistical analysis to check if the observed spot is consistent with a texture. We conclude that it is, and propose further tests. If confirmed, the detection will provide a unique window onto physics at extremely high energies.
  

Texture unwinding events may be approximated by a spherically symmetric scaling solution ({\it 21\/}), 
on a comoving radius $r < \kappa \tau$ where $\kappa$ is a fraction of unity and $\tau$ is the conformal time when the texture unwinds.  Such events lead to hot and cold spots, with a fractional temperature distortion 
\begin{equation}
\frac{\Delta T}{T}(\theta) =\pm \epsilon \frac{1}{\sqrt{1 + 4 ( \frac{\theta}{\theta_C} ) ^2}},
\label{eq:texture}
\end{equation}
where $\theta$ is the angle from the center, the amplitude is set by $\epsilon= 8 \pi^2 G \phi_0^2$ and the scale parameter $\theta_C \equiv 2 \sqrt{2} \kappa (1+z)/ \left(E(z) \int_0^z \frac{d\bar{z}}{E(\bar{z})}\right)$ and $E(z)=\sqrt{\Omega_M (1+z)^3 +\Omega_\Lambda}$, with $\Omega_M$ and $\Omega_\Lambda$ the present day matter and dark energy density parameters and $z$ the redshift of the unwinding texture. Since the scaling profile is not valid at large comoving radii $r$, we truncate Equation(\ref{eq:texture}) beyond its half-maximum by matching its value and derivative to a Gaussian.
A scale-invariant distribution of spots on the sky is predicted ({\it 29\/}), with the number of spots of scale $\theta_C$ or above, $N_{spot}(>\theta_C) = 4 \pi \nu \kappa^3 /(3\theta_C^2)$. Here $\nu$ parameterizes the comoving number density $n$ of unwinding textures:  $d n/d \tau = \nu \tau^{-4}$. 
High-resolution simulations of $SU(2)$ textures yield $\kappa \approx 0.1$ and $\nu \approx 2$ ({\it 30\/}). The corresponding fraction of the sky covered by textures with $\theta_C$ larger than 1$^\circ$ is $f_S \approx 0.017$.

We perform a Bayesian analysis of the probability ratio $\rho$ of two hypotheses given the data (see e.g. (24)). The null hypothesis 
$H_0$ describes the data $\mathbf{D}$ as a statistically homogeneous and isotropic Gaussian random field (CMB) plus instrument noise.
The alternative hypothesis $H_1$ describes the data as CMB plus noise and an additional template, $\mathbf{T}$,  given by a cosmic
texture with parameters $\epsilon$ and $\theta_C$. The evidence is the average of the likelihood $L$ with respect to the prior, $\Pi$: $E_{i} = Pr(\mathbf{D}|H_i) = \int{ L_i(\mathbf{\Theta}_i|H_i)\Pi(\mathbf{\Theta}_i)d\mathbf{\Theta}_i }$, where the $\mathbf{\Theta}_i$ are the parameters in hypothesis $H_i$. This formula naturally incorporates an Occam factor favoring the hypothesis with fewer parameters. Our template fitting is performed in a circular area of $20^\circ$ radius centered on $b = -57^\circ, l = 209^\circ$, 
in the three--year foreground cleaned WMAP combined-frequency map (WCM) ({\it 11\/}) at $\approx 1^\circ$ resolution ({\it 29\/}). 

The posterior probability ratio $\rho= Pr(H_1|\mathbf{D})/Pr(H_0|\mathbf{D})= E_1 Pr(H_1)/(E_0 Pr(H_1))$ can be used to decide between the hypotheses. The alternative hypothesis is favored when $\rho > 1$ and rejected otherwise. The a priori probability ratio for the two models, $Pr(H_1)/Pr(H_0)$, is usually set to unity, but 
since we center the template at an a posteriori selected pixel,
we set $Pr(H_1)/Pr(H_0)$ to the fraction of sky covered  by textures.
To compute $E_1/E_0$, we need the likelihood and normalized priors. The likelihood function is just $L \propto \exp{\left(-\chi^2/2\right)}$, where $\chi^2 = (\mathbf{D}-\mathbf{T})^T \mathbf{N}^{-1} (\mathbf{D} - \mathbf{T})$
%
%
and $\mathbf{N}$ is the generalized noise matrix including CMB and instrument noise ({\it 29\/}). 

As a conservative prior on $\epsilon$, we choose $0 \leq \epsilon \leq 10^{-4}$, the latter value being the upper limit imposed by the large scale Cosmic Background Explorer (COBE) satellite measurements ({\it 23--25\/}).  The prior for $\theta_C$ is obtained by normalizing the number of spots distribution, $dN_{spot}/d\theta_C \propto \theta_C^{-3}$ to unity between $\theta_{min}$ and $\theta_{max}$. Photon diffusion would smear out textures smaller than a degree or so, hence we set $\theta_{min} = 1^\circ$. At large scales textures are unlikely because the sky is finite: we set $\theta_{max} = 15^\circ$. 

We find the probability ratio $\rho \approx 2.5$, favoring the texture plus Gaussian CMB over the Gaussian-only model. The data, the best fit template and their difference are presented in Figure~(1). The best fit amplitude and scale are $\epsilon = 7.7\times 10^{-5}$ and $\theta_C = 5.1 ^\circ$. Marginalizing the posterior (i.e likelihood times prior, see Figure~(2)), we find $\theta_C = {4.9^\circ}_{-2.4}^{+2.8}$ and $\epsilon = 7.3_{-3.6}^{+2.5}\times 10^{-5}$ at $95\%$ confidence.
The value of $\epsilon$ inferred in this way from a single extreme event is biased by the detection of signals with high noise, i.e.,  large Gaussian fluctuations. To check this we generated 500 all-sky, Gaussian CMB simulations ({\it 10,29\/}) and added one cold texture spot to each, with amplitude $\epsilon = 4\times 10^{-5}$ below the upper limit, $5\times 10^{-5}$ inferred from the observed CMB anisotropy spectrum ({\it 26\/}). We perform the same template fit we applied to the data on each cold spot and then select the spots with high posterior probability ratios, $\rho > 1$. The mean amplitude obtained from these spots is $\epsilon  \approx 7.9 \times 10^{-5}$, hence there is significant overestimation. Moreover, a more realistic model of textures would predict some dispersion in the spot strength, with stronger spots caused by asymmetric, multiple, or moving textures. Again, estimating $\epsilon$ from the strongest texture spot would lead to a biased value.

As a complementary test for the a posteriori selection of the template centre, we should also study whether prominent Gaussian CMB spots produce such high values of $\rho$. Following the same procedure for 10,000 Gaussian simulations we select the most prominent spot, finding that these spots show typical values 
of $\rho \approx 0.14 < 1$, with only $\sim 5.8\%$ of the simulations showing spots with $\rho > 2.5$. Since the kurtosis of the data shows a more significant departure from Gaussianity, the percentage may further decrease if spots of all sizes were taken into account. 

In order to analyse further the CMB signal from textures, we generate 10,000 texture plus Gaussian CMB and noise simulations, and repeat the analyses performed on Gaussian simulations with no textures. Considering that we observe one $5.1^\circ$ texture in about $40\%$ of the sky of the WCM -- the region unaffected by the Galaxy -- we predict around $68$ hot and cold spots above $\theta_{min} = 1^\circ$. We generate $68$ spots per simulation with $N_{spot}(>\theta_C) \propto \theta_C^{-2}$ and $\theta_{min} = 1^\circ$, assigning a random sign and position to each spot on the sky. The amplitude is set to $\epsilon = 4\times 10^{-5}$ following the discussion above. We then repeat the previously performed multiscale analysis of skewness and kurtosis ({\it 10,11\/}). 
As there are on average the same number of hot and cold spots, the skewness is little affected. However, on the contrary, the kurtosis is increased so the anomalously high kurtosis of the data at scales around $5^\circ$ is actually compatible with the Gaussian CMB plus textures interpretation (see Figure~3).

From the results of the texture simulations, the predicted number of spots $N_{spot}(>\theta_C) \approx 1.1$ for $\theta_C\approx 5.1^\circ$, consistent with the single observed spot.  Hence we find that the abundance, shape, size and amplitude of the spot are consistent with the texture interpretation. The symmetry breaking scale corresponding to the inferred amplitude is $\phi_0 \approx 8.7 \times 10^{15} GeV$. From the relations given below Equation (\ref{eq:texture}), the observed texture unwound at $z \sim 6$, after the reionization of the intergalactic medium and potentially within reach of very deep galaxy or quasar surveys.

Further observations could test the texture hypothesis. First, if the spot is due to a texture, it was caused by a time-dependent gravitational potential. There would be no associated CMB polarization. However, if the spot is a rare statistical fluctuation in the primordial density, a correlated polarization signal, namely a preference for a radial pattern of CMB polarization around it, is expected. On these scales, for adiabatic perturbations with standard recombination, almost half the polarization signal is correlated with the temperature anisotropy ({\it 27\/}). 

Secondly, there should be many smaller texture spots, with the scale-invariant distribution described above. These would be masked or confused by the background Gaussian signal where it has maximal power, at $\theta \sim 1^\circ-2^\circ$. Nevertheless, each spot would deviate from the expected polarization-temperature correlation and a combined all-sky measurement might show a significant difference from the standard prediction. 

Finally, a texture at $z\sim 6$ would gravitationally lens objects behind it with a lensing angle of order $\epsilon$ radians. In particular, it would lens the second-order CMB anisotropies (the Vishniac effect) which peak at these angular scales. 

While certainly radical, the texture hypothesis is the most plausible explanation yet proposed for the spot. Our analysis shows it to be favored over a purely Gaussian CMB. Alternate explanations, such as voids with radius around $200$ h$^{-1}$ Mpc, are far more radical and seem inconsistent both with standard cosmology and with galaxy survey observations ({\it 28\/}).

\begin{quote}
{\bf References and Notes}

\begin{enumerate}

\item Bennett C.L., et al., 2003, ApJS, 148, 1.

\item Hinshaw G. et al., 2007,  ApJS, 170, 288.

\item Spergel D. et al., 2007, ApJS, 170, 377.

\item de Oliveira-Costa A., Tegmark M., 2006, Phys. Rev. D 74, 023005.

\item Copi C. J., Huterer D., Schwarz D. J., Starkman G. D., 2007,  Phys. Rev. D 75, 023507. 

\item Land K., Magueijo J., 2007, MNRAS.378, 153. 

\item McEwen J.D., Hobson M.P., Lasenby A.N., Mortlock D.J., 2006, MNRAS, 371, 50.

\item Eriksen H. K.,  Banday A. J., G\'{o}rski K. M., Hansen F. K., Lilje P. B., 2007, ApJ, 660, 81.

\item Wiaux Y., Vielva P., Mart\'{\i}nez-Gonz\'alez E., Vandergheynst P., 2006, Phys Rev. Lett., 96, 1303.

\item Cruz M., Cay\'on L., Mart\'{\i}nez--Gonz\'alez E., Vielva P., Jin J., 2007, ApJ, 655, 11. 

\item Vielva P., Mart\'{\i}nez--Gonz\'alez E., Barreiro R. B., Sanz J.L., Cay\'on L., 2004, ApJ, 609, 22. 

\item Cruz M., Mart\'{\i}nez--Gonz\'alez E., Vielva P., Cay\'on L., 2005, MNRAS, 356, 29. 

\item Cruz M., Tucci M., Mart\'{\i}nez--Gonz\'alez E., Vielva P., 2006,  MNRAS, 369, 57. 

\item Tomita K., 2005, Phys. Rev. D, 72, 10.

\item Inoue K. T., Silk J., 2007, ApJ 664, 650.%

\item Jaffe T. R., Hervik S., Banday A. J., G\'{o}rski K. M., 2006, ApJ 644, 701. 

\item Bridges M.,  McEwen J. D, Lasenby A. N., Hobson M. P., 2007, MNRAS, 377, 1473. 

\item Turok N., 1989, Phys. Rev. Lett., 63, 2625.

\item Turok N. \& Spergel D. N., 1990, Phys. Rev. Letters, 64, 2736.

\item Kibble T.W.B., 1976, J. Phys., {\bf A9} 1387. 

\item Vilenkin A., Shellard E.P.S., {\it Cosmic strings and other topological defects}, Cambridge University Press, Cambridge, 2000.

\item Hobson M.P., McLachlan C., 2003, MNRAS, 338, 765.

\item Pen U., Spergel D. N., Turok N., 1994, Phys. Rev. D, 49, 692.

\item Bennett D.P., Rhie S-H., 1993,  Astrophys.J. {\bf 406} L7.

\item Durrer R., Kunz M., Melchiorri A., 1999, Phys.\ Rev.\  D {\bf 59} 123005.
%
\item Bevis N., Hindmarsh M., Kunz M., 2004, Phys. Rev. {\bf D70} 043508. 
\item Crittenden R. G., Coulson D., Turok N., 1995, Phys. Rev. {\bf D52} 5402.

\item Hoyle F., Vogeley M.S., 2004, ApJ, 607, 751.
\item Supporting online material is available on Science OnlineSupporting. 

\item See http://www.damtp.cam.ac.uk/cosmos/viz/movies/neil.html.

\item MC thanks the Ministerio de Educacion y Ciencia for a predoctoral FPU fellowship. NT thanks STFC (UK) and the Centre for Theoretical Cosmology in Cambridge for support. PV thanks a I3P contract from the CSIC. MC, PV and EMG thank the Ministerio de Educacion y  Ciencia, ref.ESP2004-07067-C03-01. The authors acknowledge the use of the Legacy Archive for Microwave Background Data Analysis (LAMBDA). Support for LAMBDA is provided by the NASA Office of Space Science. We acknowledge the use of the HEALPix software.

\end{enumerate}
\end{quote}


\bibliography{scibib}

\bibliographystyle{Science}




\clearpage

\begin{figure*}
\begin{center}
\includegraphics[height=150mm]{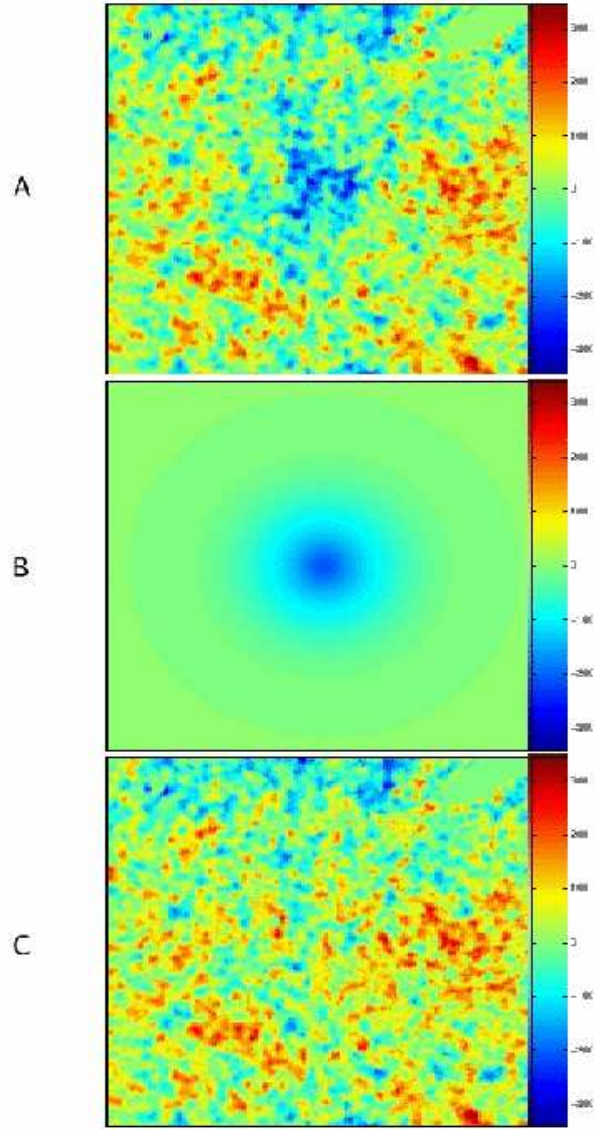}
\end{center}
\caption{A: Azimuthal projection of a $43^\circ \times 43^\circ$ patch of the WCM, centered at ($b = -57^\circ, l = 209^\circ$). 
B: Best fit texture template. C: WCM subtracting the texture template.
The temperature units shown in the colorbars are $\mu K$ and the pixel is $13.7$ arcmin. The y-axis points to the Galactic north pole. (The template is available on request).}
\end{figure*}


\begin{figure*}
\begin{center}
\includegraphics[height=150mm]{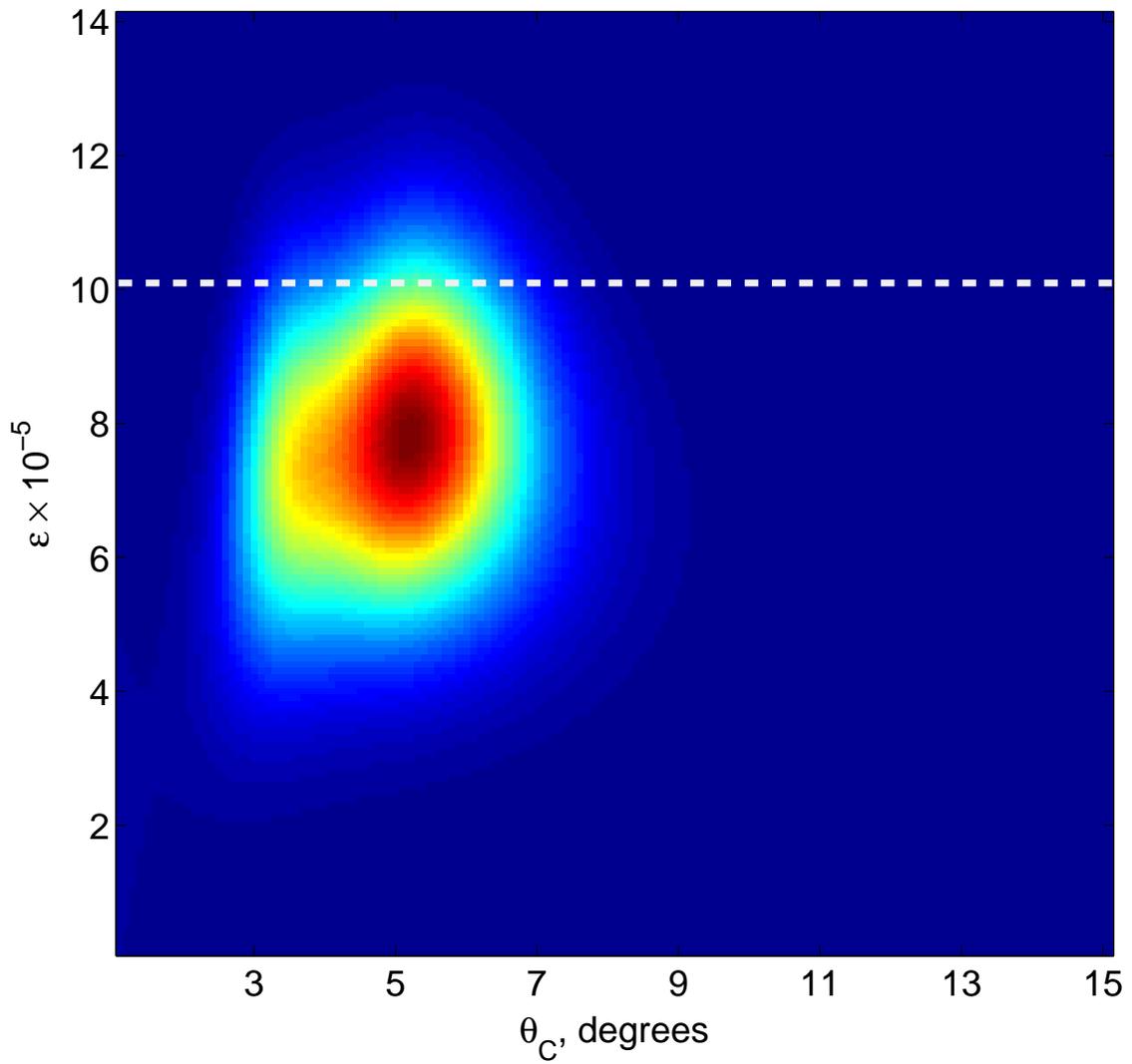}
\end{center}
\caption{Posterior (likelihood times prior) as a function of amplitude $\epsilon$ and scale $\theta_C$ for the texture template fit. The prior limit on the amplitude is marked by a dashed white line.}
\end{figure*}


\begin{figure*}
\begin{center}
\includegraphics[height=150mm]{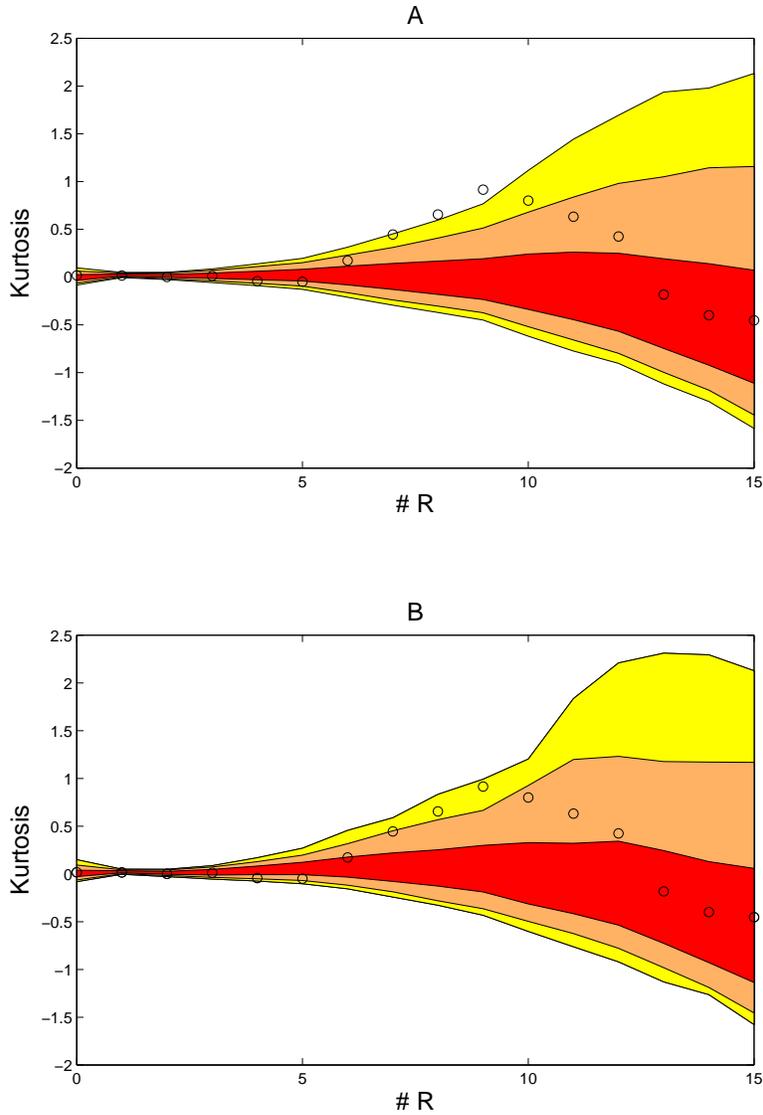}
\end{center}
\caption{Kurtosis of the WCM convolved with the SMHW at 15 scales. The bands represent the $68\%$ (red), $95\%$ (orange) and $99\%$ (yellow) acceptance intervals given by 10,000 simulations of Gaussian CMB (A) and Gaussian CMB plus textures (B). 
The WCM data (circles) show deviation from the expected values compared to Gaussian simulations, but are fully consistent with the Gaussian CMB plus textures interpretation at all scales.
$\# R$ stands for scale number. The 15 scales are: $R_1 = 13.7$, $R_2 = 25$, $R_3 = 50$, $R_4 = 75$, $R_5 = 100$, $R_6 = 150$, $R_7 = 200$, $R_8 = 250$, $R_9 = 300$, 
$R_{10} = 400$, $R_{11} = 500$, $R_{12} = 600$, $R_{13} = 750$, $R_{14} = 900$ and $R_{15} = 1050$ arcmin.}
\end{figure*}


\clearpage
\newpage

\section*{Supporting online material}

\subsection*{Supporting Text}

\subsubsection*{Data and Simulations}

We use the foreground-cleaned WMAP combined map (WCM) of the three year WMAP data release ({\it S1\/}). This map is a linear combination of the three maps at frequencies where the CMB is the dominant signal, namely 41, 61 and 94 GHz. 
The kp0 mask is applied to the data in order to exclude foreground contaminated pixels. This mask is the most conservative one 
of those proposed by the WMAP team and excludes around $23\%$ of the sky including a few hundreds of known point sources.
In the wavelet analyses a set of extended masks ({\it S2,S3\/}) is used in order to exclude those pixels convolved with
pixels in the mask. The extension is performed around the Galactic part of the mask and depends on the wavelet scale $R$.
Here we use an extension of $2.5R$ although the results are quite independent of the exact extension ({\it S2\/}). Hence for $R=5^\circ$ the mask excludes about $60\%$ of the sky.

In the template fitting part, we use the WCM in the HEALPix pixelization scheme ({\it S4\/}) with resolution
parameter Nside = 64. Since the scale of the cold spot we are interested in, is around $5^\circ$ (diameter of $\sim 10^\circ$) this resolution is good enough
and reduces the number of pixels used in the template fitting. 

Gaussian simulations for each frequency channel are produced, based on the best fit power spectrum given by the WMAP team ({\it S1\/}). After convolving with the adecquate beam, uncorrelated Gaussian noise realizations are added, taking into account the number of observations per pixel. The maps are then combined as for the data.
The resolution parameter used in the wavelet skewness and kurtosis analyses, is Nside = 256.

\subsubsection*{Template fitting}

We perform a template fitting in a circular area of $20^\circ$ radius centered at Galactic coordinates ($b = -57^\circ, l = 209^\circ$) in order to check whether there is a texture-like template in the WCM at the position of the spot. 

Excluding from the analysis the point sources masked in the three year kp0 mask, the total number of pixels considered is 1438.
Although the angular size of the cold spot is about $10^\circ$ we have to consider at least a $20^\circ$ radius
patch to take into account the whole neighborhood for the fit. 
The Spherical Mexican Hat Wavelet (SMHW) convolves all the pixels in this region and they could contribute in an important way to 
the detected structure.

To compute the generalised noise matrix of the template fitting, $\mathbf{N}$, which describes the correlations
between all the pixels, the CMB and the noise contributions have to be worked out.
The calculation of the latter is straightforward since the number of observations per pixel  and hence the noise variance is known. As the instrumental noise is uncorrelated the noise matrix is diagonal 
In order to obtain the CMB contribution, we calculate the correlation function for the WCM taking into account the pixel and beam effects.
As a complementary test we calculate the CMB correlation matrix through 70000 Gaussian simulations. 
Comparing the resulting differences of logarithmic evidences obtained from the WCM correlation function with those 
obtained using simulations, the errors are below $1\%$.

Choosing different extrapolations of the temperature profile of Equation (2), as an exponential or a SMHW, 
we obtain $\Delta\mbox{ln} E$ values between 4.7 and 5.2, which still give $\rho>1$. 

\subsubsection*{Number of textures}

The expected number of hot and cold spots due to textures, with an angular size larger than some $\theta_C$ (measured in radians), is given by  
\begin{equation} 
N_{spot} = \int d\tau {dn \over d \tau} 4 \pi (\tau_0-\tau)^2 \int_{\theta_C (\tau_0-\tau)}^{\kappa \tau} 2 dr, 
\end{equation}
where the number of texture unwindings per comoving volume, $dn$, per conformal time, $\tau$, is $(d n / d \tau) = \nu \tau^{-4}$, with $\nu$ a dimensionless constant $\nu \approx 2$ as measured in numerical simulations. Here, $\tau_0$ is the present conformal time, $\kappa$ a fraction of unity and the factor $4 \pi (\tau_0-\tau)^2$ is the comoving area of the sphere of currently detected CMB photons at the conformal time $\tau$ when the texture unwinds. If the unwinding event is inside the sphere, the photons ``fall in" to an outgoing spherical shell of stress-energy and a blue spot is produced. Whereas if the event is outside the sphere, the photons ``climb out" of the ingoing shell and a red spot is produced. The upper limit $r <\kappa \tau$ is imposed to account for the finite size of the region described by the single-texture scaling solution, and the factor two accounts for hot and cold spots. The angular scale subtended by the resulting hot or cold spots is given (in the small angle approximation, and assuming a spatially flat universe) by $\theta= r/(\tau_0-\tau)$. If we consider spots larger than some size $\theta_C$, this imposes the lower limit $r> \theta_C (\tau_0-\tau)$ and the $r$ integral is nonzero only for conformal times $\tau > \tau_0 \theta_C /(\kappa+\theta_C)$.

We shall be interested in the regime where $\theta_C \lsim \kappa$, in which case the integral simplifies to
\begin{eqnarray}
N_{spot} \approx {4 \pi \over 3} \nu {\kappa^3 \over \theta_C^2}.
\end{eqnarray}
The number of hot and cold spots of angular radius between $\theta_C$ and $\theta_C +d \theta_C$ is just the differential of the previous Equation, namely 
\begin{equation}
\frac{dN_{spot}}{d \theta_C} = {8 \pi \nu \over 3} \frac{\kappa^3}{ \theta_C^3},
\end{equation}
which corresponds to Equation (2) in the main text.
It follows that the expected fraction of the sky covered by spots of angular radius greater than $\theta_C$ is
\be
f_S = {\langle A \rangle \over 4 \pi} \approx \int_{\theta_C}^1 d \theta {\pi \theta^2} {2 \nu \kappa^3 \over 3\theta^3 } = {2 \pi \nu \kappa^3 \over 3} {\rm ln} (1/\theta_{C}),
\label{b3}
\ee
where we approximated the upper limit, where the small angle approximation breaks down, as unity. Setting $\nu=2$, $\kappa=0.1$ and $\theta_C=1^\circ$ we obtain $f_S=0.017$.

\subsubsection*{Theories giving rise to texture}

As mentioned in the article, texture is produced whenever a simple Lie group such as $SU(2)$ is completely broken. However, there are two important cases to distinguish. The conceptually simpler case is that of a global symmetry, in which case the texture evolves as described in the article. The order parameter 
is initially random, with a microscopic correlation length, but it becomes aligned on progressively larger scales, with the correlation length growing at a fixed fraction of the speed of light. However, if the symmetry is gauged, then the gauge field can simply relax to cancel spatial gradients of the order parameter (called a Higgs field in this case), so that the texture quickly relaxes to a zero energy density state. The texture winding number can still be important -- for example, in the standard electroweak theory, it corresponds to the baryon number of the universe -- but it does not continue to evolve in the late universe. 

It is sometimes argued on general theoretical grounds that the only fundamental symmetries of nature should be gauged, not global. For example, it is believed that string theory would never predict an exact global symmetry. However, global symmetries are perfectly consistent with quantum field theory and many of the simplest theories known do produce texture, even if they do not possess any scalar fields. As an example, consider $N$ massless Dirac fermions coupled to a Yang-Mills field with a nonabelian gauge group G. This theory possesses a global $SU(N)_L\times SU(N)_R$ chiral symmetry. The full symmetry would be unbroken at high temperature in the deconfining phase, but would break to the diagonal subgroup $SU(N)$ as the system cooled below the confining transition, leading to the formation of texture. One interesting example of a hypothetical global symmetry is family symmetry, for which a natural candidate is $SU(3)$ since there are three families of elementary particles. In this case, the $SU(3)$ cannot be gauged because then it would be anomalous ({\it S5\/}). Finally, there are known mechanisms in which a theory which has only gauge symmetry at a fundamental level can, with exponentially tiny corrections, lead to an effective global symmetry ({\it S6\/}). Therefore, while the general theoretical arguments are certainly important, they are not conclusive. Of course, this makes the potential observation of a texture even more interesting and significant.

\subsubsection*{Collapse and unwinding processes: Nonlinear sigma model}

As the universe expands, the gradient energy in the symmetry-breaking field (or order parameter) is red-shifted away: the initial, random configuration evolves in a scaling manner where the field progressively orders itself on a scale set by the Hubble horizon. If the vacuum manifold possesses a nontrivial topology (for texture, a nontrivial third homotopy group $\pi_3$), then in some regions of space there will be a topological obstruction to field ordering. In such regions, the only way for ordering to occur is for the winding configuration to draw itself together and collapse down to a microscopic scale, so that the field gradients become strong enough to pull the field off the vacuum manifold and over the potential energy barrier. 

The process of field ordering may be described with excellent accuracy by the nonlinear sigma model (NLSM) ({\it S7\/}). For the simplest case of $SU(2)$ texture, the space of vacua is a three-sphere and the appropriate NLSM is just the $O(N)$ vector model with $N=4$, with the order parameter being a four-component vector $\vec{\phi}$ whose length is fixed: $\vec{\phi}^2=\phi_0^2$. The equation of motion in an expanding universe is 
\begin{equation}
\ddot{\vec{\phi}} + 2 {\dot{a}\over a} \dot{\phi} -\nabla^2 \vec{\phi}  = \phi_0^{-2} \left((\nabla{\vec{\phi}})^2 -\dot{\vec{\phi}}^2\right) \vec{\phi},
\end{equation}
and this is the equation we solve numerically. Since the textures of interest all unwound in the matter-dominated era, long before the dark energy dominated, we use a flat FRW metric with scale factor $a \propto \tau^2$. 

The numerical algorithm used to evolve the texture was described in detail in ({\it S8}), with one minor detail. The ``spin flip'' procedure described there was, as noted by Borrill et al. ({\it S9}), unnecessary so it is not used. The code has been parallelized for runs of up to $1024^3$ on the COSMOS supercomputer and is made publicly available on the website \hfill\break 
http://www.damtp.cam.ac.uk/cosmos/viz/movies/neil.html. In the new simulations textures as small as $0.1$ of the horizon are now well-resolved, yielding a considerably higher value of $\nu$ and lower value of $\kappa$ compared to early simulations conducted over a decade and a half ago ({\it S10}).


\newpage

\subsection*{Supporting References}

\begin{itemize}

\item[S1.]
Hinshaw et al., 2007,  ApJS, 170, 288.

\item[S2.]
Vielva P., Mart\'{\i}nez--Gonz\'alez E., Barreiro R. B., Sanz J.L., Cay\'on L., 2004, ApJ, 609, 22. 

\item[S3.]
Cruz M., Mart\'{\i}nez--Gonz\'alez E., Vielva P., Cay\'on L., 2005, MNRAS, 356, 29. 

\item[S4.]
G\'{o}rski K.M., Hivon E. F., Wandelt B. D., Banday J., Hansen F. K., Barthelmann M., 2005, ApJ 622, 759.

\item[S5.]
Joyce M., Turok N., 1994, Nucl. Phys. B416, 389.

\item[S6.]
Turok N., 1996, Phys.\ Rev.\ Lett.\  76, 1015.
%

\item[S7.]
Turok N., 1989, Phys. Rev. Lett., 63, 2625.

\item[S8.]
Pen U., Spergel D. N., Turok N., 1994, Phys. Rev. D, 49, 692.

\item[S9.]
Borrill J., Copeland E.J., Liddle A.R., Stebbins A. and Veeraraghavan S., 1994,  Phys.Rev. D 50, 2469.

\item[S10.]
Spergel D. N., Turok N. G., Press, William H., Ryden B. S., 1991, Phys. Rev. D, 43, 1038.

\end{itemize}

\end{document}

\end{document}